\documentclass[12pt,preprint]{aastex}

\usepackage{natbib}

\slugcomment{}

\shorttitle{Deep Radio Upper Limits on Ultracool Dwarfs}
\shortauthors{Osten et al.}

\newcommand{\bd}[1]{Luhman\,16AB}

\begin{document}
\title{The Deepest Constraints on Radio and X-ray Magnetic Activity in Ultracool Dwarfs from WISE~J104915.57-531906.1  }
\shorttitle{No Magnetic Activity in Luhman\;16AB}
 
\author{Rachel A. Osten\altaffilmark{1}}
\affil{Space Telescope Science Institute}
\affil{3700 San Martin Drive, Baltimore, MD 21218}
\altaffiltext{1}{Also at Center for Astrophysical Sciences, Johns Hopkins University, Baltimore, MD 21218}
\email{osten@stsci.edu}

\author{Carl Melis}
\affil{University of California, San Diego}

\author{Beate Stelzer}
\affil{INAF }
\affil{Osservatorio Astronomico di Palermo}
\affil{Piazza del Parlamento 1, 90134 Palermo, Italy}

\author{Keith W. Bannister\altaffilmark{2}}
\affil{CSIRO Astronomy and Space Science}
\affil{PO Box 76, Epping NSW 1710, Australia}
\altaffiltext{2}{Bolton Fellow}

\author{ Jackie Radigan}
\affil{Space Telescope Science Institute}

\author{Adam J. Burgasser}
\affil{University of California, San Diego}

\author{Alex Wolszczan\altaffilmark{3}, Kevin L. Luhman\altaffilmark{3} }
\affil{Department of Astronomy and Astrophysics, The Pennsylvania State University, University Park, PA 16802, USA}
\altaffiltext{3}{Center for Exoplanets and Habitable Worlds, The Pennsylvania State University, University Park, PA 16802, USA}

\begin{abstract}
We report upper limits to the radio and X-ray
emission from the newly discovered ultracool dwarf
binary WISE~J104915.57$-$531906.1 (\bd{}). 
As the nearest ultracool dwarf binary (2 pc), its proximity offers
a hefty advantage to studying plasma processes in ultracool dwarfs which are more similar in gross
properties (radius, mass, temperature) to the solar system giant planets than stars.
The radio and X-ray emission
upper limits from the Australia Telescope Compact Array (ATCA) and Chandra observations, each spanning multiple rotation periods,
provide the deepest fractional radio and X-ray luminosities to date on an ultracool dwarf,
with
$\log{(L_{\rm r,\nu}/L_{\rm bol}) [Hz^{-1}]} < -18.1$ (5.5 GHz),
$\log{(L_{\rm r,\nu}/L_{\rm bol}) [Hz^{-1}]} < -17.9$ (9~GHz), 
and $\log{(L_{\rm x}/L_{\rm bol})} < -5.7$. 
While the radio upper limits alone do not allow for a constraint on the magnetic field strength, we
limit the size of any coherently emitting region in our line of sight to less than 0.2\% of the radius of one of the brown dwarfs.
Any source of incoherent emission must span less than about 20\% of the brown dwarf radius, assuming magnetic field strengths
of a few tens to a few hundred Gauss. 
The fast rotation and large amplitude photometric variability exhibited by the T dwarf
in the \bd{} system are not accompanied by enhanced nonthermal
radio emission, nor enhanced heating to coronal temperatures, as observed on some higher
mass ultracool dwarfs,
confirming the expected decoupling of matter and magnetic field in 
cool neutral atmospheres.
\end{abstract}


\section{Introduction }
Recently, the discovery of a brown dwarf binary only 2 parsecs from the Sun was announced, 
making it the third closest system after the Alpha-Centauri system and Barnard's star \citep{luhman13}.
With L7.5 and T0.5 spectral types \citep{burgasser13} the \bd{} system
(also known as WISE~J104915.57$-$531906.1)
has quickly become a benchmark for the study of ultracool atmospheres.  
A unique feature of the \bd{} system is the large amplitude photometric variability \citep[11\% in an $i+z$ 
filter with a period of 4.87$\pm$0.01\,hr;][]{gillon13} of the T0.5 component, with rapid evolution of
the global weather patterns on timescales of about a day \citep{crossfield2014}.
The two components of \bd{} are separated by 1.5", or 3 AU \citep{luhman13}, so not able to influence
each other via magnetic interactions.



In principle, photospheric features causing photometric variability could be magnetic- or cloud-related; while there
have been sporadic measurements of magnetic activity in mid-L and later spectral type dwarfs,
no clear trends have emerged which connect photometrically variable and magnetically active ultracool dwarfs.
Since the 
atmospheres of ultracool dwarfs are increasingly neutral, they are less likely to support cool magnetic spots \citep{gelino02,mohanty02} 
than their earlier type stellar counterparts, and the observed variability in late L- and T-dwarfs 
has been attributed to the presence of patchy 
clouds \citep{ackerman01,burgasser02,marley10}.  
Despite this, there is evidence that in at least some cases, photometric variability of 
ultracool dwarfs is linked to magnetic activity
\citep{clarke03,lane2007,harding2013} 

Magnetic activity signatures in ultracool dwarfs are rare: to date, only 5 L dwarfs and one T dwarf have been detected in the radio band, 
and only 1 L dwarf and no T dwarf has been detected in the X-ray band, with a wide range of behaviors displayed among
the small number of detections.
Studies have shown that for late-M 
to early-L dwarfs faster rotation results in an increased radio detection fraction 
\citep{mclean2012} while X-ray emission seems to be suppressed leading to a sort 
of super-saturation \citep{Berger10.1}.

Dynamo models explain the generation of magnetic fields in ultracool dwarfs by
extrapolating convection-driven geodynamo models with strong density stratification \citep{christensen2009}:
part of the convected energy flux is converted to magnetic energy to balance ohmic diffusion.
Such scaling laws predict quite strong magnetic fields, of order 1 kG
for a 1~GY old, 0.05 M$_{\odot}$ brown dwarf with T$_{\rm eff}=$1500 K and average density of 90,000 kg m$^{-3}$.
These scalings do not, however explain why only a handful of  ultracool dwarfs of spectral type L and T have been detected through 
radio observations (implying field strengths compatible with these extrapolations) while other 
objects have considerably lower upper limits. 
Other parameters must govern the generation of radio emission and/or
field strength.


Due to the 
proximity of the \bd{} system, its magnetic activity can be probed with unprecedented sensitivity.  An absence of 
activity signatures would support the prevailing view that large amplitude 
photometric variability of early T-dwarfs is not connected to magnetism, 
but rather is a consequence of patchy cloud coverage.
We report on two epochs of radio observations of \bd{}
with the Australia Telescope Compact Array in March and
May 2013 and on a Chandra X-ray pointing carried out in November 2013
\footnote{We are also aware of observations of 
\bd{}
with the South African KAT-7 array at a wavelength of 20 cm, which
took place over 7 hours on April 8, 2013 (R. Fender, private communication).  The rms image noise is 4 mJy and
 there is no source near the expected position of the brown dwarf binary.  Because of the 
factor of $\sim$1000 disparity in upper limit of the decimeter wavelength observations compared with the centimeter wavelength
upper limits, we concentrate on the ATCA results in the following discussion.}.
These observations provide the most sensitive constraints
to date on the radio and X-ray emission from ultracool dwarfs.

\section{Observations }

\subsection{Radio Observations}

\begin{table}[hbtp]
\begin{center}
\caption{Upper Limits from Radio Observations \label{tbl:radio}}
\begin{tabular}{llllll}
Frequency & Epoch  & t$_{\rm int}$& $\sigma$ & beamsize & pos. angle\\
 GHz & YYYY-MM-DD & hrs &  $\mu$Jy & "$\times$" & $^{\circ}$ E of N\\
\hline
5.5 & 2013-03-09 & 9 & 6.5  & 2.3$\times$1.2  &7\\
" & 2013-05-02 & 6.5 & 8.5  &  2.7$\times$1.3 &$-$28\\
" & both epochs & 15.5 & 5 &2.1$\times$1.3 & $-$5\\
9& 2013-03-09 & 9& 7.5  & 1.4$\times$0.8  &7\\
" & 2013-05-02 & 6.5 & 15  & 2.0$\times$0.8  &$-$19\\
" & both epochs & 15.5 & 6.8  & 1.4$\times$0.8  &3\\
\hline
\end{tabular}
\end{center}
\end{table}

\begin{figure}
\begin{center}
\includegraphics[scale=0.5]{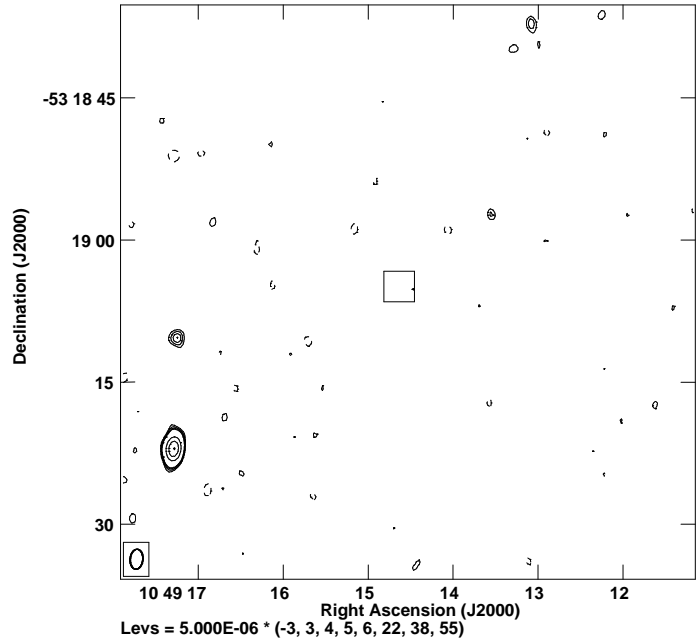}
\caption{Plot of the 5.5 GHz image around the expected position of \bd{} from combining both epochs
of observations. 
The box at the center has a length of 3.2", corresponding to a combination of position uncertainties;
see text for details. The beam is shown in the lower left of the image.
\label{fig:radsky}}
\end{center}
\end{figure}

\bd{} was observed twice with the
ATCA: in the 6A configuration (baselines of 0.337--5.94\,km) 
on 09 March 2013 and again with the 6C configuration 
(baselines of 0.153--6.0\,km) on 02 May 2013 (UT). 
Continuum mode observations were taken
on both dates in dual-sideband mode simultaneously at 5.5\,GHz  and 9.0\,GHz. 
The Compact Array Broadband Backend \citep[CABB][]{wilson2011} was used with
2\,GHz bandwidth per observing frequency and 2048 channels each 1\,MHz wide.
The gain calibrator QSO\,B1036$-$52 was used for both epochs, with 
primary flux calibrator  QSO\,B1934$-$638; \bd{} was tracked in 10-minute intervals for both epochs.
The flux calibrator was used also as the bandpass calibrator during the March observations,
but RFI at early times during the May observations prevented it from also being used
as a bandpass calibrator; instead 
a single scan of QSO\,B1036$-$52 taken at high
elevation was used for bandpass calibration.
All data were reduced using the AIPS package 
\citep{aips} and best practices for wide-band data reduction.
Table~\ref{tbl:radio} lists the beam sizes for each epoch and frequency band.

Data at 5.5 and 9 GHz for each epoch were imaged separately and after being 
combined into a single data set per band; both Stokes I and V were searched
for 
\bd{}. 
We propagated the $WISE$
position at epoch MJD 55380.018731 by the proper motions given in \citet{luhman13} to 
get the expected position coordinates 
(J2000~10 49 15.99 $-$53 19 04.9 for March and 10 49 16.01 $-$53 19 04.9 for May).
There is only a 0.25" difference in the position between the two radio epochs. 
The error range for the expected position of \bd{} is comprised of $\pm$0.75" from the
binary separation, an upper limit of $\pm$0.5" from parallactic motions (likely smaller due to
the $\sim$ two month separation of the epochs), $\pm$0.07" from position uncertainty 
\citep[taken from][]{luhman13}, and $\pm$0.02"
from propagating proper motion uncertainties stated in \citet{luhman13}. 
No source was found near 
the expected position for the brown dwarf pair in either epoch. 
Figure~\ref{fig:radsky}
shows the 5.5 GHz radio sky around this expected position, with a box of $\pm$1.6" encompassing the 
maximum of all the errors stated above.
The nearest statistically detected source is 12.5" away, with a flux density at 5.5 GHz of 30 $\mu$Jy.
No bursty emission in Stokes I or V 
is evident in either of the bands in either epoch, in 
light curves with bin sizes of 60, 300, and 600~seconds 
\citep[see e.g.,][]{ostenwolk2009}.
Details of the observations (and sensitivities derived from individual and combined epochs) are listed in Table~\ref{tbl:radio}.
A 1$\sigma$ upper limit of 5 (6.8) $\mu$Jy/beam at an observing frequency of 5.5 (9.0) GHz, and at a distance of 2pc, translates into a
3$\sigma$ radio luminosity upper limit
$L_{\nu}$ of 7.2$\times$10$^{10}$  (9.8$\times$10$^{10}$) erg s$^{-1}$ Hz$^{-1}$
for \bd{}. 

%

\subsection{X-ray observations}

Chandra observed \bd{} on 10 Nov 2013 for $50$\,ks (ObsID 15705)
using the ACIS-S3 detector.
The data analysis was performed with the CIAO software 
package\footnote{CIAO is made available by the CXC and can be downloaded 
from \\ http://cxc.harvard.edu/ciao/download/} version 4.6. 
The analysis started with the level\,1 events file provided by the
{\em Chandra} X-ray Center (CXC). 
In order to optimize the spatial resolution, pixel randomization was removed.
The events file was filtered on event grades
(retaining the standard grades $0$, $2$, $3$, $4$, and $6$), 
and the standard good time interval file was used. 

\begin{figure}
\includegraphics[scale=0.7]{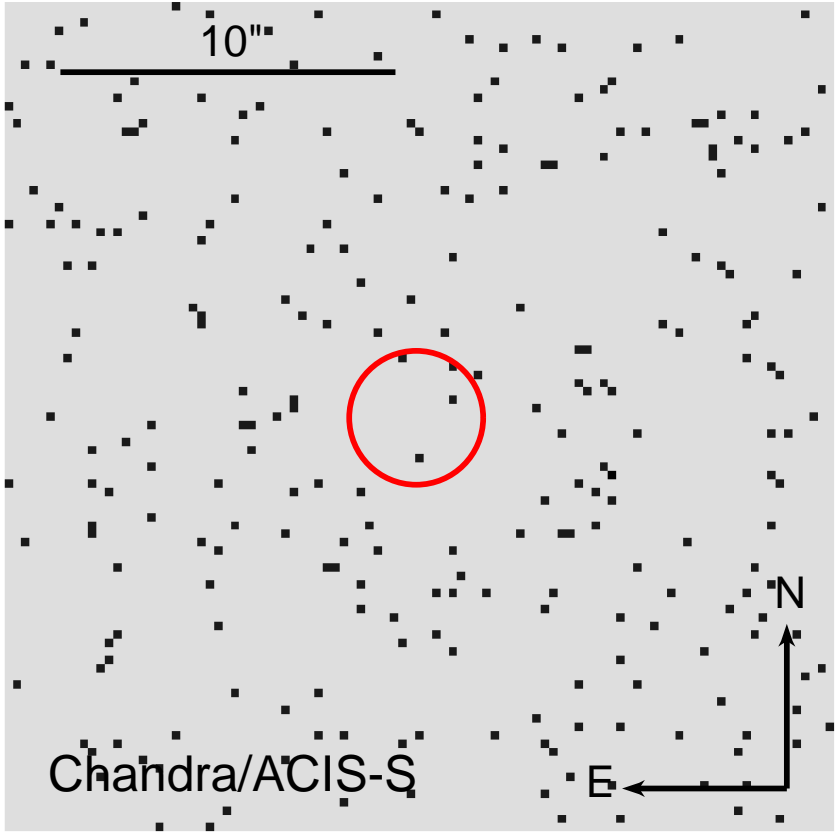}
\caption{Figure showing the region of the X-ray sky around the expected position of \bd{}
at the time of the Chandra observation.  Red circle indicates the 90\% encircled energy radius of 2" centered
at the expected position of \bd{}; as discussed in Section 2.2, this encompasses the maximum deviation
from the expected position by propagating errors in position, proper motion, parallax, and binary separation.
\label{fig:xraysky}
}
\end{figure}

We determined the position of \bd{} at the epoch of the {\em Chandra} observations
using the proper motion given by \cite{luhman13}. The predicted coordinates
are J2000 10 49 14.65 -53 19 05.2, and there is no evidence a of source at this location
(see Figure~\ref{fig:xraysky}).
As discussed in Section 2.1, the uncertainty in the
expected position is 
a maximum of 1.34". 
For an on-axis source, 90\% of the encircled energy lies within a radius of 2" 
(Chandra Proposer's Observatory Guide), and we used this as our guide for 
establishing the spatial region in which to probe for any X-ray emission. 
The
closest detected X-ray source is separated by $57^{\prime\prime}$ from this position,
with a total of 17 counts.

Calculation of an upper limit proceeded with estimation of the background rate at the
position of \bd{}.
90\% of the encircled energy lies within 2$^{\prime\prime}$ of the central pixel at an energy
of 1.49 keV (Chandra Proposer's Observatory Guide).
Using an annulus extending from 2$^{\prime\prime}-10^{\prime\prime}$ around the position of \bd{},
we calculate a mean background rate within 2$^{\prime\prime}$
of the target, or 0.37 counts (0.2-2 keV) in 48.35 ks. 
We calculated the quantile distribution for a Poisson distribution with this intensity in the R statistical
computing software package \citet{rcite},
and find an upper limit of 2 counts at a significance level of P=0.001; this corresponds
to a confidence level of 99.9\%, 
equivalent to a Gaussian sigma level of 3.09 \citep{gehrels1986}. 
For the on-source exposure time of $48.35$\,ks, the upper limit count rate is then
$4.1 \cdot 10^{-5}$\,cts/s. 
We calculated the count-to-flux conversion factor for a one-temperature thermal 
plasma (APEC model) with PIMMS\footnote{http://cxc.harvard.edu/toolkit/pimms.jsp}
and we verified that it is insensitive for reasonable assumptions on the plasma temperature
considering the negligible absorption expected for the $2$\,pc distance of \bd{}. 
We thus constrain the X-ray luminosity in the 0.2-2 keV band to $\log{L_{\rm x}}\,{\rm [erg/s]} < 23.0$. 

\section{Discussion}

\subsection{Magnetic Activity Constraints for \bd{}}

\begin{figure}
\includegraphics[scale=0.6,angle=90]{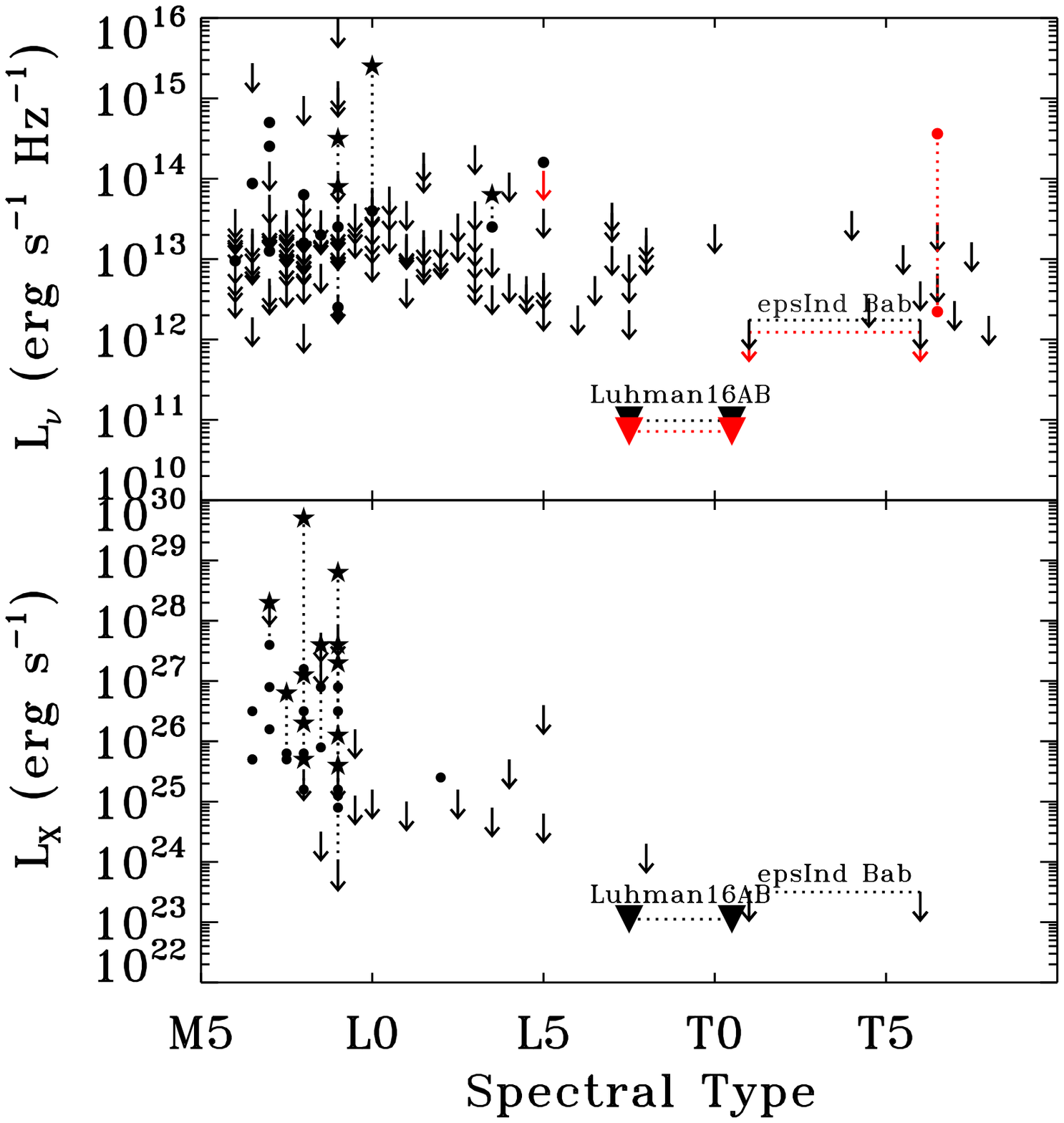}
\caption{\textbf{(Top)} Radio luminosity versus spectral type for ultracool dwarfs with spectral types 
of M6 and later. 
The bulk of the data refers to the $8$\,GHz band and is shown in black; $5$\,GHz observations are shown in red.
Data are taken from  the compilations of
\citet{mclean2012}, \citet{williams2014},
\citet{burgasser2013}, \citet{route2012} and \citet{williams2013}.
Downward-pointing arrows are 3$\sigma$ upper limits, while filled circles correspond to detections.
Dotted vertical lines connect measurements of the same object;
a star symbol connects flare measurements with measurements or limits on quiescence of the same object
at the same wavelength.
The radio upper limits for
the \bd{} system are assigned to either the L7.5 
or T0.5 component of the system and connected by horizontal lines. 
\textbf{(Bottom)} X-ray luminosity versus spectral type for stellar objects with spectral types 
of M6 and later. Data are taken from \citet{williams2014},
\citet{cook2014}, 
\citet{audard2005}, and \citet{stelzer2012}.
Symbols are as in the top panel.
\label{fig:rxplot}
}
\end{figure}

The upper limits of \bd{} presented in this work are the 
strongest constraints obtained so far for the radio and X-ray luminosity of any 
ultracool dwarf.
We compute the radio and X-ray activity indices, 
$\log{(L_{\rm r,\nu}/L_{\rm bol})}$ and $\log{(L_{\rm x}/L_{\rm bol})}$, 
making use of the bolometric luminosities given by \cite{Faherty2014} for both components
of the binary. 
We evaluate the activity indices separately for the L7.5 and the T0.5 component,  
assuming that only one of the binary components is possibly magnetically active. 
However, the bolometric luminosities of Luhman\,16A and 16B are almost the same 
and the error we make by using their average is likely smaller than the
sum of all other uncertainties. 
We find $\log{(L_{\rm r,\nu}/L_{\rm bol})} < -18.1$ (5.5 GHz),
$\log{(L_{\rm r,\nu}/L_{\rm bol})} < -17.9$ (9 GHz), 
and $\log{(L_{\rm x}/L_{\rm bol})} < -5.7$. 
Figure~\ref{fig:rxplot} puts these upper limits in the context of detections
and upper limits for other ultracool dwarfs.

The upper limits at 5.5 and 9 GHz for \bd{}
from ATCA are a factor of 15 and 10, respectively, lower than the most sensitive upper limit of any other ultracool
dwarf. 
The X-ray upper limit is about a factor of two deeper than previous sensitive upper limits.
$\epsilon$~Ind~Bab (T1+T6) is the next closest brown dwarf with sensitive limits on 
X-ray and radio emission, from \citet{audard2005}. At a distance of 3.6 pc, it is only a factor of 1.8
further away than \bd{}, conveying about a factor of three difference in luminosity sensitivity, and 
measurements were made for both systems with the same radio and X-ray facilities: ATCA and 
the {\it Chandra} X-ray Observatory. 
The disparity in radio upper limits can largely be attributed to the increase in bandwidth available with the CABB
on ATCA now (2 GHz) compared with what was available for \citet{audard2005}'s observation (128 MHz).

\subsection{Interpretation}

We have presented the most sensitive upper limits on X-ray and radio emission for ultracool dwarfs to date. 
The T dwarf component has a measured rotation period of $4.87$\,h \citep{gillon13}.
Our X-ray limit for \bd{} 
confirms for late L and T dwarfs
the previous evidence gained from M/L dwarfs for a sharp drop of X-ray activity levels despite 
fast rotation. 
This absence of X-ray 
activity is most likely associated with the high electrical resistivities
in such cool atmospheres which prevent the coupling of matter and magnetic field which
is necessary to develop magnetic activity \citep{mohanty02}. 
\citet{fleming1995} detected stellar coronal heating efficiencies (as measured by $L_{X}$/L$_{\rm bol}$)
down to levels approximately a factor of five lower than our upper limit. 
Our upper limit is also consistent with
the activity levels of the Sun at the highest points of its activity cycle, which reaches a maximum
of $\log{L_{X}/L_{\rm bol}}=-5.9$ in the 0.2-2.4 keV band
\citep{peres2000}.

As Figure~\ref{fig:rxplot} shows, there is a marked drop-off in the number of radio 
detections for objects later than mid-L, with a range of radio 
luminosities observed at a fixed spectral type.
Detections of radio emission in ultracool dwarfs are often used to argue for the 
existence of strong magnetic fields. However, the inverse is not true:
the lack of radio detection does not allow for a determination of magnetic field strengths
in either of these objects, contrary to the statements made in \citet{berger2006} for upper 
limits on radio emission to a sample of ultracool dwarfs. 
Important conclusions regarding the physical extent of any 
emission can be drawn from the radio flux density upper limits in
examination of the conditions under which these mechanisms operate.

The interpretation of the variable radio emission in ultracool dwarfs 
has centered around the action of an electron-cyclotron maser operating in a region of high 
magnetic field strength \citep{nichols2012}. In this scenario, the observing frequency is tied
to the electron-cyclotron frequency in the emitting region and is related to 
$\nu_{c}=2.8\times10^{6}$ B (MHz) by $\nu_{\rm obs}=s\nu_{c}$ for harmonic number $s$ equal to 1 or 2, implying
kG fields in the radio-emitting region detected at cm wavelengths, and consistent with extrapolations
from convection-driven geodynamo scaling laws \citep{christensen2009}.
The intensity of radio emission expected from a coherent process 
such as this is not predictable based solely
on the number of emitting particles or magnetic field strength.
However, the high brightness temperatures required for coherent emission 
\citep[usually taken to be T$_{b}>$10$^{12}$K;][]{coherent}
coupled with the 
radio flux density upper limit and observing frequency sets a stringent upper limit on 
the size scale of any radio-emitting region. 
Rewriting the standard equation \citep{dulk1985} for parameters applicable to the current 
case, and taking the dwarf radius to be approximately 1 Jupiter radius, in line with
measurements \citep{bdradius},
leads to the following constraints:\\
\begin{equation}
x_{R_{J}} = 1316 \frac{d_{pc}}{\nu_{\rm GHz}} \left( \frac{S_{\rm \mu Jy}}{T_{\rm b}} \right)^{1/2}
\end{equation}
where $x$ is the size of any radio-emitting region in units of $R_{J}$ ($R_{J}$ is one Jupiter radius (=7.1$\times$10$^{9}$ cm)),
$d_{\rm pc}$ is the distance to the dwarf in pc, $\nu_{GHz}$ is the observing frequency in GHz,
S$_{\rm \mu Jy}$ is the flux density in $\mu$Jy, and $T_{b}$ is the brightness temperature in K. 
Evaluated for the upper limits at the two frequencies (and assuming T$_{b}=$10$^{12}$K) gives $x \le$0.002 at 5.5 GHz, and
$x \le$0.001 at 9 GHz.  
These are upper limits, as stellar phenomena have demonstrated the existence of
brightness temperatures as high as T$_{b} \approx$10$^{18}$K \citep{ostenbastian2008}.

The upper limit on size holds if the conditions are right to produce coherent emission. 
Growth rates of the cyclotron maser instability are maximized in relatively rarefied, magnetized plasmas 
where the dimensionless ratio of the plasma frequency to the electron-cyclotron frequency
is less than a few \citep{lee2013}.
The atmosphere calculations of \cite{mohanty02} showed that in the lower atmosphere, the total density
for a dusty atmosphere model in a cool dwarf with T$_{\rm eff}$ near 1500 K will be about 10$^{-9}$ g cm$^{-3}$,
with an ionization fraction of about 10$^{-11}$. This would suggest an electron density of approximately
10$^{3}$ cm$^{-3}$.
The kG field strengths for these objects derived from scaling laws, combined with these parameters, indicate
that the conditions for the instability may exist,
but the cyclotron maser mechanism could be inoperable for reasons still to be determined.
\citet{mutel2007} found a strong dependence of the growth rates of the cyclotron maser instability
on the opening angle of the loss cone distribution of electrons that could power the instability.
Beaming effects may also explain the lack of detections, if there is a misalignment between the opening angle of the
emission and the line of sight.  

\begin{figure}
\includegraphics[angle=90,scale=0.5]{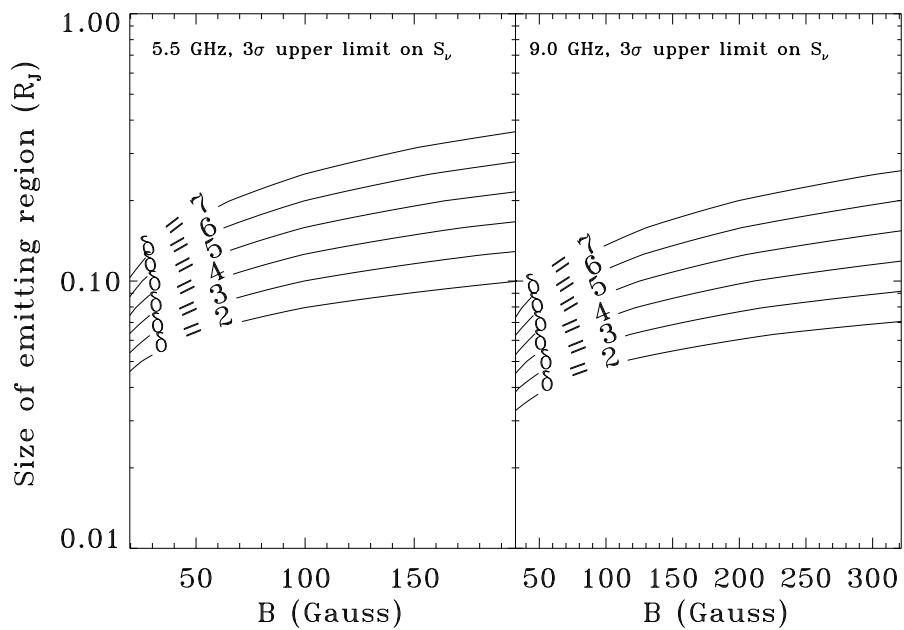}
\caption{Constraints on the size of the emitting region and magnetic field strength from the upper limits on the radio
flux density at 5.5 and 9.0 GHz of \bd{}. These constraints are calculated assuming that one of the dwarfs is capable of
producing gyrosynchrotron emission from a power-law distribution of electrons in a high magnetic field region.
The size of the emitting region is given in units of Jupiter radii, $R_{J}$. Each curve gives the upper limit of the region of
parameter space allowed for the specified value of $\delta$, the index of the distribution of accelerated particles
producing the gyrosynchrotron emission.
Anything below the line is compatible with the upper limits
for that combination of parameters.
\label{fig:limits}
}
\end{figure}

Another possibility 
that has been put forward to explain the quiescent radio emission from 
ultracool dwarfs 
is gyrosynchrotron emission, in analogy with the magnetic activity
seen in higher mass dwarf stars \citep{gudel2002}.
For this incoherent process, the strength of the emission depends not only on the magnetic field
strength in the radio-emitting source, but also on 
the index of the distribution of accelerated particles with energy ($\delta$), and the size of the emitting region.
Figure~\ref{fig:limits} displays the values of $\delta$, $B$, and size of the emitting source that are compatible with
the observed upper limit on flux density at the two radio frequencies, given the limit of applicability of the analytic
expressions in \citet{dulk1985}.
While the constraints on size are not as stringent as for the case of a coherent emission, they do 
rule out a global gyrosynchrotron-emitting magnetosphere around one of the dwarfs in the \bd{} system, as this would lead
to detectable levels of gyrosynchrotron emission.

Our discussion has concentrated mainly on steady-levels of emission. 
 The observation of radio bursts in a relatively small sample of all radio-observed
UCDs combined with the small rotation period of the known radio bursters 
has given rise to the discussion of selection effects, e.g., the typical length
of the radio observations (few hours) may not have covered the -- generally unknown -- 
full rotational cycle of many UCDs \citep{stelzer2012}. This bias can be ruled out here.  
Multiple rotation periods of the T0.5 dwarf were covered with our radio data, 
so the observations were sensitive to bursts occurring at particular rotation phases.
Yet, the viewing geometry and/or topology of the magnetic field may prevent the
detection of such bursts $-$ if present $-$ on \bd{}.
Limits presented here are likely unachievable for other ultracool dwarfs in the near future,
only to be exceeded possibly by measurements from the Athena mission (for X-rays) and
the Square Kilometer Array or Next Generation Very Large Array future radio telescopes.


\acknowledgements
RAO acknowledges support from the Chandra X-ray Observatory under grant GO4-15140X.
C.M. acknowledges support from the National Science Foundation under
award No.\ AST-1003318. 
The Center for Exoplanets and Habitable Worlds is supported by the Pennsylvania State University, the
Eberly College of Science, and the Pennsylvania Space Grant Consortium.
Thanks to Mark Wieringa for his help with the March ATCA observations and data analysis.
Thanks also to Eric Feigelson for help in setting up the first ATCA observation, and for
substantive comments on the statistics of upper limits.



\end{document}